\documentclass[%
 reprint,
nofootinbib,
 amsmath,amssymb,
 aps,
prd,
]{revtex4-2}
\usepackage{color}
\usepackage{orcidlink}
\usepackage{hyperref} 
\usepackage{cleveref} 

\hypersetup{colorlinks = true, 
	    linkcolor = blue, 
	    urlcolor = blue,
            citecolor = blue} 

\usepackage{graphicx}
\usepackage{dcolumn}
\usepackage{bm}

\begin{document}

\preprint{APS/123-QED}

\title{High-energy neutrino emission from tidal disruption event outflow-cloud interactions}

\author{Hanji Wu~\orcidlink{0000-0003-3003-866X}}
\affiliation{
Department of Astronomy, School of Physics and Technology, Wuhan University, Wuhan 430072, China\\
}
\author{Kai Wang~\orcidlink{0000-0003-4976-4098}}%
\affiliation{%
Department of Astronomy, School of Physics, Huazhong University of Science and Technology, Wuhan 430074, China\\
}%
\author{Wei Wang~\orcidlink{0000-0003-3901-8403}}
\affiliation{
Department of Astronomy, School of Physics and Technology, Wuhan University, Wuhan 430072, China\\
}%



\date{\today}

\begin{abstract}

Tidal disruption events (TDEs), characterized by their luminous transients and high-velocity outflows, have emerged as plausible sources of high-energy neutrinos contributing to the diffuse neutrino. In this study, we calculate the contribution of TDEs to the diffuse neutrino by employing the outflow-cloud model within the TDE framework. Our analysis indicates that the contribution of TDEs becomes negligible when the redshift $Z$ exceeds 2. Employing a set of fiducial values, which includes outflow energy $E_{\rm kin}=10^{51}$ erg, a proton spectrum cutoff energy $E_{\rm p,max}=100$ PeV, a volume TDE rate $\dot{N}=8 \times 10^{-7}\ \rm Mpc^{-3}\ year^{-1}$, covering fraction of clouds $C_V=0.1$, energy conversion efficiency in the shock $\eta =0.1$, and a proton spectrum index $\Gamma=-1.7$, we find that TDEs can account for approximately 80\% of the contribution at energies around 0.3 PeV. Additionally, TDEs still contribute around 18\% to the IceCube data below 0.1 PeV and the total contribution is $\sim 24^{+2}_{-15}\%$. In addition, we also discuss the potential influence of various parameter values on the results in detail. With the IceCube data, we impose constraints on the combination of the physical parameters, i.e., $C_{f}=\dot{N}E_{\rm kin}C_{\rm v}\eta$. Future observations or theoretical considerations would fix some physical parameters, which will help to constrain some individual parameters of TDEs.


\end{abstract}

\maketitle


\section{introduction}\label{intro}
Tidal disruption events (TDEs) represent a class of transient phenomena wherein a supermassive black hole (SMBH) at the center of a galaxy disrupts a nearby star, unleashing a significant burst of energy. It's important to note that for a TDE to occur, the star must remain outside the event horizon, imposing an upper limit on the black hole mass—roughly below $1 \times 10^{8}M_{\odot}$ for nonspinning (Schwarzschild) black holes and around $7 \times 10^{8}M_{\odot}$ for maximally spinning (Kerr) black holes. Beyond these limits, the event horizon would exceed the tidal disruption radius \citep{beloborodov1992angular,kesden2012black}. The aftermath of a TDE involves the star's debris falling back onto the SMBH, resulting in luminous outbursts typically observed in optical/ultraviolet or X-ray bands \citep{komossa2015tidal}. These outbursts carry substantial energy, not only in the form of electromagnetic radiation \citep{hills1975possible,lidskii1979tidal,rees1988tidal,evans1989tidal,phinney1989manifestations,ulmer1999flares} but also through ultra-fast outflows, directly confirmed via UV and X-ray observations \citep{blanchard2017ps16dtm,blagorodnova2019broad,hung2019discovery,nicholl2020outflow}. These outflows are a product of relativistic apsidal precession; as the stream of star debris collides with still-falling debris after passing the pericenter, collision-induced outflows are generated \citep{lu2020self}. Additionally, the fallen debris triggers the accretion disc to enter a high accretion mode, launching energetic outflows \citep{curd2019grrmhd}. Numerical research indicates that these outflows can attain velocities of approximately $0.1c$ and the energy of outflows even reaching $\sim 10^{52}\rm erg$ \citep{lu2020self,curd2019grrmhd}. The presence of outflows is further evidenced by radio emissions \citep{alexander2016discovery,alexander2017radio,stein2021tidal}, which have been detected in TDE candidates, emitting energy in the range of $10^{36}-10^{42}\rm erg\ s^{-1}$ over periods ranging from days to years \citep{alexander2020radio}. These radio emissions not only support the existence of outflows but also constrain their kinetic luminosity to a range of $10^{43}-10^{45}\rm erg\ s^{-1}$ \citep{stein2021tidal,cendes2021radio,mou2022radio, kara2018ultrafast}. Moreover, outflows persist for several months, whether due to violent self-interaction or super-Eddington accretion \citep{mou2021years}. Given their significant kinetic energy, estimated at $10^{50}-10^{52}\rm erg$ when considering the duration and kinetic luminosity, outflows are consistent with numerical predictions.

The outflow propagates and interacts with the material around a SMBH. In relatively high luminosity active galactic nuclei (AGNs), there is generally a broad-line region (BLR, the distance of 0.01-1 pc from the SMBH, due to the Keplerian velocities) including many dense clouds (density about $10^{10}\rm cm^{-3}$ \citep{osterbrock2006astrophysics}) around the SMBH \citep{khachikian1974atlas,barth1996ultraviolet,zhang2007size,ho2008nuclear,netzer2015revisiting}. Even in lower-luminosity AGNs, hidden BLRs have been inferred from deep Keck spectropolarimetric observations \citep{barth1999polarized1,barth1999polarized2,tran2001hidden}, although, they didn't have the broad lines in the spectrum. As for the quiescent galaxy, there is not enough flux to illuminate the environment around the center SMBH, so it is a challenge to determine if there are dense clouds at $\sim 0.01-1\rm pc$. Furthermore, the model that estimates the magnetic field and energy of cosmic ray electrons from synchrotron emission in radio band to limit the shock energy and outflow kinetic energy provided by \citep{duran2013radius}, and that is used in the known radio TDEs \citep{matsumoto2021radio} favor the clouds with a much higher density than the Sgr $\rm A_{\ast}$ profile \citep{mou2022radio}. In this paper, we assume that there are dense clouds around the SMBHs. When the outflows impact the clouds, the bow shock will be generated on the cloud exteriors, potentially accelerating charged particles \citep{drury1983introduction} reaching $\sim\rm PeV$ via diffusive shock acceleration (DSA) processes. Notably, secondary particles produced by hadronic interactions involving high-energy protons play a crucial role in multi-messenger astronomy \citep{kachelriess2019cosmic,meszaros2019multi,olinto2021poemma}. 

Due to the development of IceCube, a large-volume Cherenkov detector \citep{achterberg2006first}, the IceCube-Gen2 \citep{ishihara2023next} which is made of $8\,\rm km^{3}$ optical array with the effective area $100\rm m^{2}$ at around $\sim \rm PeV$ \citep{aartsen2019search}, has allowed for the exploration of the origins of high-energy neutrinos. Notably, a significant event, IC-170922A, with an energy of approximately $290\rm TeV$, was associated with the blazar TXS 0506+056 on September 22, 2017 \citep{icecube2018neutrino} and the most significant neutrino point source is at the coordinates of the type 2 Seyfert galaxy NGC 1068 \citep{aartsen2020time} (interestingly, although, the type 2 Seyfert galaxies only contain narrow lines, the broad emission line also could be detected in the polarised spectrum from using the technique called spectro-polarimetry \citep{1989ApJ...342..224G}, particularly in NGC 1068 \citep{1985ApJ...297..621A}, which suggest that there would be the dense clouds around the center SMBH \citep{barth1999polarized1,barth1999polarized2,tran2001hidden}). This research opened a new window to high-energy neutrino astrophysics. In recent years, not only the development of IceCube but also the progress of Zwicky Transient Facility (ZTF) \citep{bellm2018zwicky}, the Palomar 48-inch Schmidt telescope which provides $8\,\rm s$ time resolution and $47\rm\ deg^{2}$ field of view, provide a method to discover the correlation of optical transients and the neutrino events \citep{stein2019search}. Up to now, the TDEs have emerged as strong candidates for the origin of the high-energy neutrino events \citep{stein2021tidal,reusch2022candidate}, including the AT2019dsg \citep{nordin2019ztf,pasham2019nicer,pasham2019swift} and AT2019fdr \citep{2019TNSTR.771....1N,2019TNSCR1016....1C} corresponding to the IceCube-191001A \citep{2019GCN.25913....1I} and the IceCube-200530A \citep{icecube2020icecube}, respectively. Thereafter, our outflow-cloud interactions model could explain well the correlation of high-energy neutrino event (IceCube-191001A) and TDE (AT2019dsg) \citep{wu2022could} in which the TDE would generate magnificent outflows that interact with the clouds and induce a bow shock to accelerate the protons and produce the high-energy neutrinos by the accelerated protons reacting with the protons in clouds. The results consist of the number of neutrinos estimated by the population bolometric energy flux \citep{stein2021tidal}. Additionally, the radio flare in $\sim\rm GHz$ by outflow-cloud model from accelerated electrons could be also produced \citep{mou2022radio}. As the outflows continue to propagate, the outflows will interact with the torus, thus the radio, X-ray, and gamma-ray afterglows from outflows-torus interaction are further investigated \citep{mou2021years,mou2021years2}. 

Due to the upgrade of the astronomical equipment, more and more TDEs can be discovered, and up to now, over 56 TDE candidates have been reported \citep{gezari2021tidal}. The increase in the number of observed TDE candidates makes the research on relatively high precision TDE rate possible. The first investigation of measuring the TDE rate from X-ray and optical survey contains the rate at $\sim 10^{-5}\rm year^{-1}\ galaxy^{-1}$ \citep{donley2002large,esquej2008evolution,gezari2009luminous,van2014measurement}, but due to the lack of TDEs in every survey, this results is an order of magnitude lower than the dynamical model expectation $\sim 10^{-4}\rm year^{-1}\ galaxy^{-1}$ \citep{magorrian1999rates,wang2004revised} which is approached by calculating the loss-cone dynamics by stellar density profiles. However, as ZTF finds more and more TDEs, \cite{van2018mass,van2020optical} estimates the rate of $\sim 10^{-4}\rm year^{-1}\ galaxy^{-1}$ that is consistent with the theoretical expectations. In this paper, we use the volume TDE rate $\dot{N}=8\times 10^{-7}\rm Mpc^{-3}\ year^{-1}$ as the fiducial value, based on the latest research \citep{van2018mass,van2020optical} on the flux average, which contains the all TDE observation until now. The TDE rate evolves as the function of redshift $Z$ is discussed in Sec. \ref{results}.


The high-energy neutrinos from TDEs are discussed by the other previous researches,  and most studies need a powerful jet to produce high energy neutrino \citep{wang2011probing,dai2017can,senno2017high,liu2020neutrino,lunardini2017high}. \cite{wang2011probing} explored the high energy protons interacting with X-ray photons by photomeson interactions. \citep{liu2020neutrino} tried to explain the coincidence between neutrino event IC-191001A and the TDE AT2019dsg with an Off-Axis jet. Furthermore, \cite{lunardini2017high} studied the high-energy neutrinos produced by TDE jet from a statistical perspective, however, the TDE with jet couldn't be dominant as neutrino sources, unless, they could have a wide-angle emission \citep{dai2017can}. Then, \citep{senno2017high} studied dark TDEs with choked jets try to explain the jetted TDEs with such low rate. There also are other works discussing the other possibilities to produce the high neutrino from TDE: \citep{murase2020high} explore the high neutrino can be produced by the corona around an accretion disk; \citep{hayasaki2019neutrino} explore the possibility of the neutrino produced by accretion disk in super-Eddington accretion phase and the radiatively inefﬁcient accretion ﬂows; besides, \citep{winter2023interpretation} studied the neutrinos propagate in different path so that there is time-delay between the TDE and the neutrino event.

In this work, we aim to calculate the neutrinos produced by the TDE outflow-clouds model which would contribute to the astrophysical diffuse neutrino. The paper is organized as follows. We briefly introduce the physical model of a single neutrino event in Sec. \ref{single}, then connect the rate to get the diffuse neutrinos in Sec. \ref{background}. In Sec. \ref{obs}, we compare with observation reported by IceCube \citep{abbasi2021icecube}, then, the results and discussion are presented in the last Section.

\section{diffuse neutrinos from outflow-cloud model}

\subsection{Physical model of single neutrino source}\label{single}
In our previous study \citep{wu2022could}, we developed an outflow-cloud interactions model to explain the intriguing correlation between the subPeV neutrino event, IceCube-191001A, and the TDE, AT2019dsg. High-energy neutrino emission could arise from the outflow-cloud interaction model. When the TDE occurs in the center of a galaxy and the outflows are produced, the outflows would interact with the clouds around the SMBH in the galaxy. Then the interaction would produce the bow shock outside of clouds, where the protons are accelerated by bow shock through the DSA mechanism. The high-energy accelerated protons would enter the cloud and interact with the protons in clouds \citep{dar1997hadronic,barkov2010gamma,barkov2012interpretation,bosch2012fermi,wang2022jet} to produce the high-energy neutrinos.

\subsubsection{the physical picture}
In the context of our previous study, a TDE occurs leading to the expulsion of outflows from the center region of the galaxy, which collide with the clouds with the covering factor $C_{\rm v}\sim 0.1$ around the SMBH and are considered as simplified spherically symmetric outflows with kinetic energy $E_{\rm kin}$. The outflows propagate and interact with clouds to produce two shocks, that is, a bow shock outside the cloud and a cloud shock sweeping through the cloud. During the outflow-cloud interaction, the kinetic energy of the outflow will be converted into the bow shock and the cloud shock. The energy ratio of the cloud shock to the bow shock is $\frac{E_{CS}}{E_{BS}}\sim \chi^{-0.5}$ with the number density ratio $\chi\equiv \frac{\rho_{c}}{\rho_{o}}$ between the cloud and the outflow~\citep{mckee1975interaction}. The energy of the cloud shock is typically much smaller than that of the bow shock since the number density of the cloud is much higher than that of the outflow, e.g., $\frac{E_{CS}}{E_{BS}}\sim$1\% for $\frac{\rho_{c}}{\rho_{o}}=$10000\footnote{The optical observation support the number density of clouds is around $\sim 10^{10}\rm cm^{-3}$ \citep{osterbrock2006astrophysics} and the dynamic of outflow implies the number density of outflow at around $\sim 10^{6}\rm cm^{-3}$. Note that if the actual solid angle $\Omega$ of outflow is lower than 4 $\pi$, it will enhance the outflow density by a factor of $4 \pi /\Omega$}\citep{mou2022radio,mou2021years,mou2021years2,wu2022could}. \citep{mou2022radio,mou2021years,mou2021years2,wu2022could}. Therefore, we neglect the cloud shock here. According to the DSA mechanism (the shock acceleration efficiency as in the other previous research \citep{bell2004turbulent,hillas2005can,malkov2001nonlinear,mou2021years2,wu2022could} as $\eta\sim 0.1$, i.e., 10\% of the shock energy converted to accelerated particles), the bow shock accelerates the proton breaking Maxwell distribution as a power-law distribution with spectral index $\Gamma$ and an exponential cutoff energy $E_{\rm p,max}$:
\begin{equation}
    \frac{dn(E_{\rm p})}{dE_{\rm p}}=K_{\rm p}E_{\rm p}^{-\Gamma}e^{-\frac{E_{\rm p}}{E_{\rm p,max}}}\label{Ep},
\end{equation}
where $\frac{dn(E_{\rm p})}{dE_{\rm p}}$ for the distribution of protons, $E_{\rm p}$ for the energy of the accelerated protons, $K_{\rm p}$ for the normalization factor to connect the kinetic energy $E_{\rm kin}$ by $C_{\rm v}\times \eta \times E_{\rm kin}=K_{\rm p}\int E_{\rm p}\frac{dn(E_{\rm p})}{dE_{\rm p}}dE_{\rm p}$. The accelerated high-energy protons following the distribution described by Eq. \ref{Ep} would react with other particles to produce neutrinos.

\subsubsection{hadronic emission}\label{hadronicemission}
There are two channels to produce neutrinos by accelerated high-energy protons: proton-proton (pp) collisions and the photomeson production (p$\gamma$) process. High-density clouds are typically located at 0.01-1 pc from central SMBH~\citep{osterbrock2006astrophysics,ho2008nuclear,netzer2015revisiting}. The distance between the clouds and the SMBH is relatively large, so the photon density is insufficient to consume the accelerated high-energy protons, namely, weak p$\gamma$ interactions. As indicated in \citep{wu2022could}, the acceleration timescale in the bow shock for a particle with the energy of $E_{\rm p}$ and the charge number of $Z$ is
\begin{equation}
    T_{\rm acc}=\frac{8}{3}\frac{c}{ZeBV_{o}^{2}}E_{\rm p},
\end{equation}
where $B$ is the magnetic field in the acceleration region of the bow shock and $V_{\rm o}$ is the velocity of TDE outflows.

\begin{figure}
    \centering
    \includegraphics[scale=0.49]{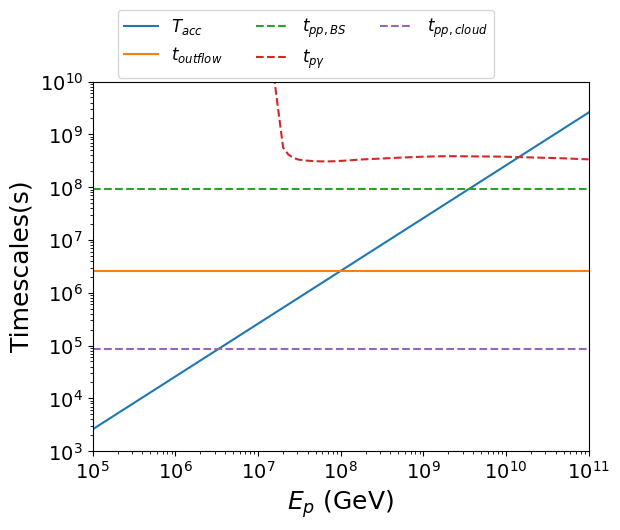}
    \caption{This figure shows the timescales of particle acceleration (blue solid line) and the pp interactions in clouds with the number density $n_{\,\rm cloud}=10^{10}\rm cm^{-3}$ (purple dash line) and in the bow shock with the number density $n_{\,\rm outflow}=10^{7}\rm cm^{-3}$ (green dash line), the p$\gamma$ interaction timescale (red dash line) according to \cite{kelner2008energy,tchernin2013exploration}, and the duration of outflow (1 month is adopted, orange solid line). Note that high-energy protons are effectively accelerated at the bow shock site, and thus the maximum proton energy is determined by $T_{\rm acc}=\min(t_{\rm outflow},t_{\rm p\gamma},t_{\rm pp,BS})$. The most dominant hadronic interaction channel is the $pp$ interactions inside the cloud. See text for more details.}
    \label{timescale}
\end{figure}

Due to the low gas density in the outflow, the time scale of the $pp$ reaction at the bow shock is large, i.e., $t_{\rm pp,BS}=(cn\sigma_{\rm pp})^{-1}\sim 3.1(\frac{n_{\rm outflow}}{10^{7}\rm cm^{-3}})^{-1}$ year with the $pp$ cross-section of $\sigma_{\rm pp}\sim 30\rm mb$, so the accelerated protons can effectively diffuse away from the bow shock. As suggested by some literature \citep{dar1997hadronic,barkov2010gamma,barkov2012interpretation,wang2022jet}, the accelerated protons can effectively reach and enter the clouds and will be consumed by the high-density gas in clouds with the timescale of $t_{\rm pp,cloud}\sim 1(\frac{n_{\rm cloud}}{10^{10}\rm cm^{-3}})^{-1}$ day. Besides, we also estimate the p$\gamma$ timescale. Here we only estimate the most optimistic p$\gamma$ reaction with the peak cross-section and the number density of TDE photons since the p$\gamma$ process is obviously insignificant. The number density of TDE photons with a typical energy $E_{\rm ph}\sim 10\,\rm eV$ at a distance of $r_o$ from the SMBH can be estimated by $n_{\rm ph}=L_{\rm ph}/(4 \pi r_o^2 c E_{\rm ph})$ with a distance of $r_o=0.01\,\rm pc$ and a TDE luminosity of $L_{\rm ph}=10^{43}\,\rm erg/s$ for AT2019dsg when the neutrino event IC-191001A occurred several months after the peak luminosity \cite{stein2021tidal}. With the typical TDE luminosity and bow shock location, the most optimistic timescale of p$\gamma$ reaction can be estimated by $t_{\rm p\gamma}\sim 3.2(\frac{n_{\rm ph}}{10^{9}\rm cm^{-3}})^{-1}$ years with the p$\gamma$ peak cross-section of $\sigma_{\rm p\gamma}\sim 0.2\rm mb$. Therefore, in our scenario, the $pp$ collisions inside the cloud dominate the hadronic interaction channel. Another key timescale is the duration of outflow (or the lifetime of the bow shock), i.e., $t_{\rm outflow}$, which is around one month. The related timescales for protons are presented in Fig.~\ref{timescale}. Protons are effectively accelerated at the bow shock site, and thus the maximum proton energy is determined by $T_{\rm acc}=\min(t_{\rm outflow},t_{\rm p\gamma},t_{\rm pp,BS})$, inducing $E_{\rm p, \max}\sim 100\,\rm PeV$. However, the maximum proton energy may deviate from this value if the actual acceleration is not in the case of Bohm diffusion~\citep{Wang_2018}, so we simply adopt $E_{\rm p, \max}$ as a free parameter below.

After acceleration in the bow shock, high-energy particles can diffuse away from the bow shock region. Motivated by some literature~\citep{dar1997hadronic,barkov2010gamma,barkov2012interpretation,wang2022jet}, we assume that a significant fraction $\alpha$ of the accelerated particles can effectively reach and enter the cloud, whereas the others will flow away bypassing the cloud. Actually, the relatively low-energy protons are likely to be carried away with the matter flow, while the accelerated high-energy protons can keep their propagation direction to some extent and tend to diffuse into the cloud. Moreover, the high-energy protons can effectively enter the cloud by suppressing the possible advection escape under certain magnetic configuration~\citep{barkov2012interpretation}. The detailed treatment of the particle diffusion is beyond the scope of this paper, therefore, a factor $\alpha<1$ is introduced by considering the uncertainty of the efficiency of cosmic rays loading into the clouds.

In addition, the protons entering the cloud can also escape from the cloud. The escape time from cloud $t_{\rm esc,cloud}$ can be evaluated by $t_{\rm esc,cloud}=\frac{r_{c}^{2}}{D_{B}}$, where $D_{B}=\frac{r_{g}^{2}\omega_{G}}{16}$ is the Bohm diffusion coefficient with the gyroradius $r_{g}=\frac{E_{p}}{eB}$ and the cyclotron frequency $\omega_{g}=\frac{eBc}{E_p}$ \citep{taylor1971plasma,kaufman1990explanation,casse2001transport,wang2022jet}, so protons need $t_{\rm esc,cloud}=4.3(\frac{r_c}{10^{14.7}{\rm cm}})^{2}(\frac{B}{1 \,\rm G}) (\frac{E_{\rm p}}{100\,\rm PeV})\,\rm days$ \footnote{In this work, the cloud shock velocity $V_{\rm cloud,shock}\sim 10^{7}\rm cm/s$ is estimated by $V_{\rm cloud,shock}=\chi^{-0.5}V_{\rm o}$ where the $V_{o}$ is the velocity of outflow (see more detail in APPENDIX A of \citep{mou2021years2}). And the magnetic field strength could be amplified to 1G (see more detail in section 2.3 of \citep{mou2022radio}).} to escape from the cloud. The escape timescale is larger than the $pp$ interaction timescale, so the escape of accelerated proton from the cloud is neglected \footnote{We assume that the particle escapes from the cloud by Bohm diffusion and Bohm limit and efficient magnetic field amplification is achieved in the cloud. Due to the uncertainty of the diffusion process, the estimation might be optimistic \citep{caprioli2012cosmic}. If the escape process is faster than the pp interactions, the neutrino flux will be suppressed by a factor of $f_{\rm pp} = \frac{t_{\rm esc,cloud}}{t_{\rm pp}}$.}.

Therefore, in our scenario, the dominant hadronic process is the $pp$ interactions inside the cloud, i.e.,
	\begin{eqnarray}
	&p+p\to p+p+a\pi^{0}+b(\pi^{+}+\pi^{-}),\\
	 &        	p+p\to p+n+\pi^{+}+a\pi^{0}+b(\pi^{+}+\pi^{-}),\label{pp2}
	\end{eqnarray}
	where $a\approx b$. The pions decay and generate $\gamma$-rays and leptons immediately:
	\begin{eqnarray}
	 &		\pi^{0}\to 2\gamma\\
    &	\pi^{+}\to \mu^{+}+\nu_{\mu},~ \mu^{+}\to e^{+}+\nu_{e}+\bar{\nu}_{\mu},\\
    &	\pi^{-}\to \mu^{-}+\bar{\nu}_{\mu},~ \mu^{-}\to e^{-}+\bar{\nu}_{e}+\nu_{\mu}. \label{n2}  
	\end{eqnarray}
These particles are the final product of pp collision in clouds, which we calculated by using the public {\rm PYTHON} package {\rm AAfragpy}\footnote{https://github.com/aafragpy/aafragpy}\citep{kachelriess2019aafrag,koldobskiy2021energy}. This calculation is based on the differential inclusive cross-section, which is from the parameterizations of QGSJET-II-04m high-energy interaction model \citep{ostapchenko2011monte,ostapchenko2013qgsjet,kachelriess2015new} and the energy of accelerated protons which is described by Eq. \ref{Ep}. 

Finally, we can get the final product spectra: 
\begin{equation}
    N_{\rm f}=\alpha\frac{dn(E_{\rm f})}{dE_{\rm f}}\label{Nnu},
\end{equation}
where $\rm f=\gamma,\nu$, etc. for the type of final particles. The typical neutrino luminosity of a single source at fiducial values is illustrated in Fig. \ref{single}.
\begin{figure}
    \centering
    \includegraphics[scale=0.49]{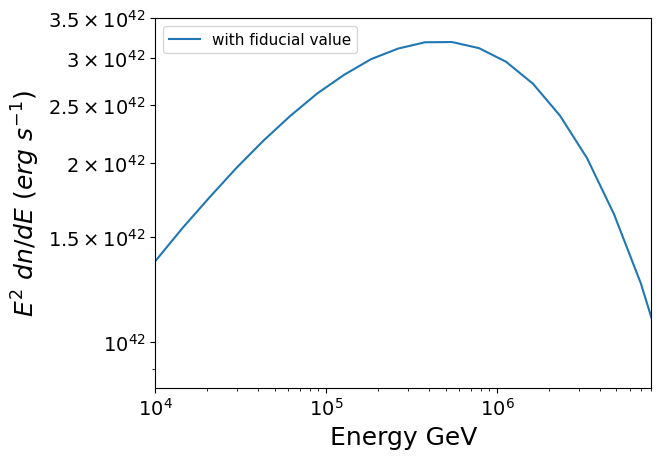}
    \caption{This figure shows the typical neutrino luminosity of a single source using the fiducial values.}
    \label{single}
\end{figure}

\begin{table*}
    \centering
    \caption{the fiducial values of model parameters}
		\label{fidu}
		\begin{tabular}{lcr} 
			\hline
			symbols & Descriptions & Fiducial Values  \\
			\hline
			$E_{\rm kin}$ & the kinetic energy of outflow & $10^{51}\rm erg$ \\
			$E_{\rm p,max}$ & the cutoff energy of protons spectrum & 100 PeV \\
			$\dot{N}$ & the TDE volume rate per cubic Mpc per year & $8\times 10^{-7}\rm Mpc^{-3}\ year^{-1}$ \\
			$\Gamma$ & the index of protons spectrum & -1.7 \\
                $C_{\rm v}$ & Covering factor of clouds & 0.1 \\
                $\eta$ & The fraction of shock energy converted to accelerated particles & 0.1 \\
                $\alpha$ & the fraction of the accelerated protons that can enter the clouds& 0.9 \\
                $B$ & Magnetic field strength & 1G \\
                null & the cosmology model & WMAP9 \\
	\hline
	\end{tabular}
\end{table*}

\subsection{Diffuse neutrino emission}\label{background}
With the in-depth study of TDE rate \citep{donley2002large,esquej2008evolution,gezari2009luminous,van2014measurement,magorrian1999rates,wang2004revised,van2018mass,van2020optical}, we could calculate the outflow-cloud interaction model contribution of the high-energy diffuse neutrino. In the latest research \citep{van2018mass,van2020optical}, the rate of TDE is estimated at $8\times 10^{-7}\rm Mpc^{-3}\ year^{-1}$ with all samples of TDEs including the most recent optical TDE discoveries. Thus, we could calculate the diffuse diffuse neutrino as follows method:

From Sec.~\ref{hadronicemission}, the single neutrino source spectrum ($N_{\rm f}=\frac{dn(E_{\rm f})}{dE_{\rm f}}$) generated by our outflow-cloud model can be calculated. Then, we assume that the volume density of TDEs is isotropic, the diffuse neutrino from our model can be calculated by:
\begin{equation}\label{neuba}
    \frac{d\dot{N}_{\rm diff}}{dE_{obs}}=\frac{1}{4\pi}\int^{\infty}_{0}\frac{N_{\nu}(E_{\nu}/(1+Z))}{4\pi D_{L}^{2}}\dot{N}4\pi r^{2}dr ,
\end{equation}
where $\frac{d\dot{N}_{\rm diff}}{dE_{\rm obs}}$ for the diffuse neutrino per unit time per unity area per steradian, $\frac{N_{\nu}(E_{\nu}/(1+Z))}{4\pi D_{\rm L}^{2}}$ for the single neutrino source flux after cosmological distance energy correction at redshift $Z$, ($N_{\nu}(E_{\nu}/(1+Z))$ is described in Eq. \ref{Nnu}), which means the single diffuse neutrino flux observed by IceCube and $D_{\rm L}$ for the luminosity distance at redshift $Z$, $\dot{N}$ for the TDE volume rate per cubic Mpc per year, $r$ for the proper distance at redshift $Z$. The last three items, $\dot{N}$, $r$, $dr$, aim to calculate TDE density on a sphere at a distance $r$ from the Earth (we assume the TDE volume density is isotropic), and we consider the accumulation of diffuse neutrino flux at different distance $r$ from 0 to $\infty$. 

Thus, several important parameters need to be set. The typical energy and velocity of TDE outflows are $10^{51}\rm erg$ and $\sim 0.1\rm c$ respectively in TDE simulation work and results \citep{curd2019grrmhd,lu2020self}, which are also confirmed by the radio flares \citep{stein2021tidal,cendes2021radio,mou2022radio,kara2018ultrafast} from TDEs reaching the radio luminosity of $\sim 10^{42}\rm erg\ s^{-1}$, then constraining the outflow kinetic luminosity $\sim 10^{44}\rm erg\ s^{-1}$ 
with velocity $\sim 0.1\rm c$ for several months. When the outflows interact with the clouds and the bow shock arises, the protons would be accelerated by the DSA mechanism:
\begin{equation}
    E_{\rm p,max}\approx\frac{3}{8}\frac{ZeB}{c}V_{o}^{2}T_{\rm acc}\label{Emax},
\end{equation}
where $B$ is the magnetic field around the bow shock with 1 G the magnetic field strength. The cutoff energy of protons spectrum $E_{\rm p,max}$ can be given by the velocity of TDE outflows $V_{\rm o}$ and the duration of the bow shock around several months which is the duration of outflows \citep{mou2021years,mou2021years2,mou2022radio} as the typical value. Thus, with the typical outflow velocity $V_{\rm o}\sim 0.1\rm c$ and the typical magnetic field strength in bow shock (1 G) and the typical timescale of acceleration (1 month), the typical cutoff energy is $100(\frac{V_{\rm o}}{0.1\rm c})^{2}(\frac{B}{1\rm G})(\frac{T_{\rm acc}}{1 month})\rm PeV$. For the index of protons spectrum, although DSA gives $\Gamma\sim -2$ in the test-particle limit \citep{drury1983introduction}, the particle feedback on the shock could produce a harder spectrum reaching $\Gamma\sim -1.5$ \citep{malkov2001nonlinear}. Due to such uncertainty of the index of protons spectrum, we chose a relatively moderate value $\Gamma= -1.7$ as the fiducial value \citep{celli2020spectral}. The TDE rate has been studied for over 20 years, and the present observation research \citep{van2018mass,van2020optical} is consistent with the theoretical research \citep{magorrian1999rates,wang2004revised}, suggesting a rate of $\sim 10^{-4}\rm year^{-1}\ galaxy^{-1}$. In this paper, we use the TDE volume rate at $\dot{N}=8\times 10^{-7}\rm Mpc^{-3}\ year^{-1}$ provided by \citep{van2018mass,van2020optical}, in which they also consider the influence of effective galaxy to TDE. Here we set the kinetic energy of outflows $E_{\rm kin}=10^{51}\rm erg$, the cutoff energy of protons spectrum $E_{\rm p,max}=100 \rm PeV$, the TDE volume rate $\dot{N}=8\times 10^{-7}\rm Mpc^{-3} year^{-1}$ and the index of protons spectrum $\Gamma=-1.7$ as fiducial values illustrated in Tab. \ref{fidu}. Besides, the larger ranges of the parameters are also tested in the results (see Fig. \ref{index}, \ref{Cf}, \ref{cutoff}) and we also discuss them in Sec. \ref{results}.

\begin{figure}
    \centering
    \includegraphics[scale=0.49]{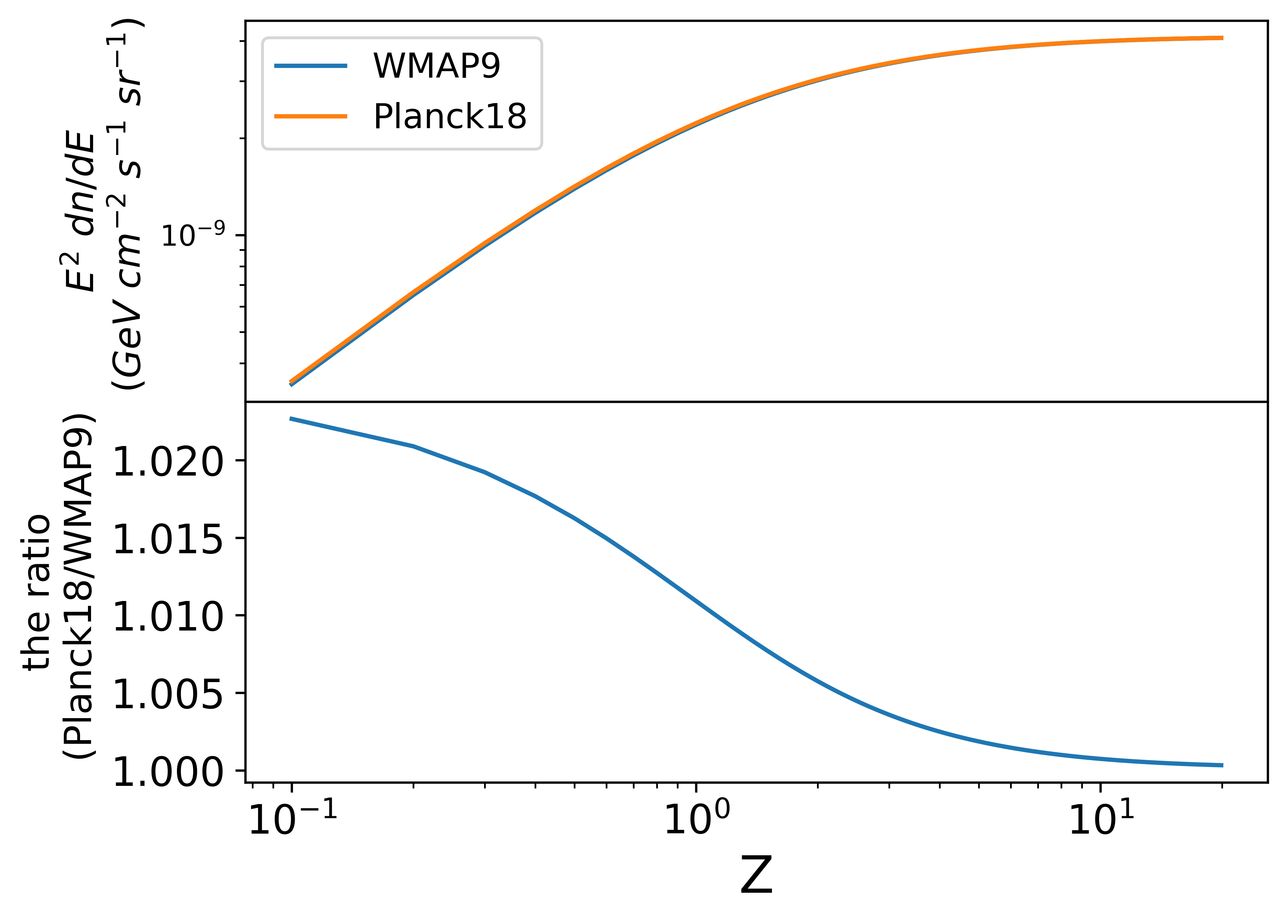}
    \caption{The top panel shows the accumulation of the diffuse neutrinos at 60 TeV from our model. Although we calculate the redshift until $Z=20$, the turning point is at about $Z\sim 2$ and the main contribution comes from the TDE at $Z<2$. The orange line represents for cosmology model Planck18, and the blue line for WMAP9. There is no distinguishing difference between them. The bottom panel shows the ratio between the neutrinos at 60 TeV produced by the outflow-cloud model from two different cosmology models Planck18 and WMAP9 as a function of redshift $Z$. At $Z>2$, the difference between these two models is less than 1\% and the most difference is about $\sim 2\%$.}
    \label{coscom}
\end{figure}

Because of the cosmological distance that we should consider, we compare the different cosmological models to limit the distance. There are two models we compared: first, WMAP9 \citep{hinshaw2013nine} which is a six-parameter $\Lambda$ cold dark matter ($\Lambda$CDM) model based on a nine-year Wilkinson Microwave Anisotropy Probe (WMAP) observation with Hubble constant $H_{0}=69.3\rm\ km\ s^{-1}\ Mpc^{-1}$, matter density $\Omega_{m}=0.28$, and cosmological constant density $\Omega_{\Lambda}=0.72$, another one, Planck18 \citep{aghanim2020planck} which is also a six-parameter $\Lambda$CDM but based on the full-mission Planck measurements of the cosmic microwave background anisotropies including the temperature and polarization maps and the lensing reconstruction with Hubble constant $H_{0}=67.7\rm\ km\ s^{-1}\ Mpc^{-1}$, matter density $\Omega_{m}=0.32$, and cosmological constant density $\Omega_{\Lambda}=0.68$. The comparing results are shown in Fig. \ref{coscom} which is the ratio of the neutrino at 60 TeV produced by the outflow-cloud model at different redshifts using two types of cosmology model results Planck18 and WMAP9. There is a relatively large difference within redshift $Z\sim 2$ which is about 1\% at redshift $Z\sim 2$ and the most difference is about 2\%. 

We also test the convergence of our model as a function of redshift $Z$ by calculating the diffuse neutrino at 60 TeV, which is shown in Fig. \ref{coscom}. At redshift $Z>2$, there is a negligible contribution, which suggests that the contribution of the reionization period and the before that is trivial, furthermore, actually, the volume TDE rate relies on the effective galaxy which is based on the M-$\sigma$ relation \citep{van2018mass,van2020optical,gezari2021tidal}, thus, it is still a debate that the uncertainty of M-$\sigma$ relation at the high redshift \citep{morishita2023enhanced}, which does not affect our model. In Fig. \ref{coscom}, the difference between the above cosmology models is negligible, so in the following part, we use the cosmology model WMAP9.

\section{results of diffuse neutrino emission from outflow-cloud model}\label{obs}
\begin{figure}
    \centering
    \includegraphics[scale=0.49]{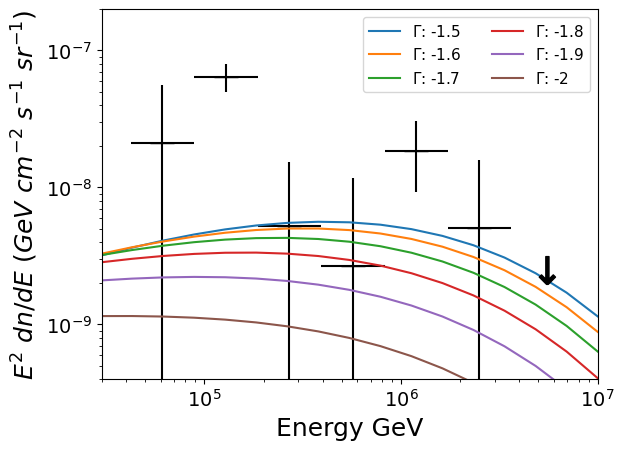}
    \caption{This figure shows the observed data from IceCube in black errorbar for data, the arrow for the upper limit, and the predicted diffuse neutrino flux from the outflow-cloud model in different indexes of the proton spectrum $\Gamma$ in -1.5 for the blue line, -1.6 for the orange line, -1.7 for the green line (the fiducial value), -1.8 for the red line, -1.9 for the purple line, and -2 for the brown line.}
    \label{index}
\end{figure}

\begin{figure}
    \centering
    \includegraphics[scale=0.49]{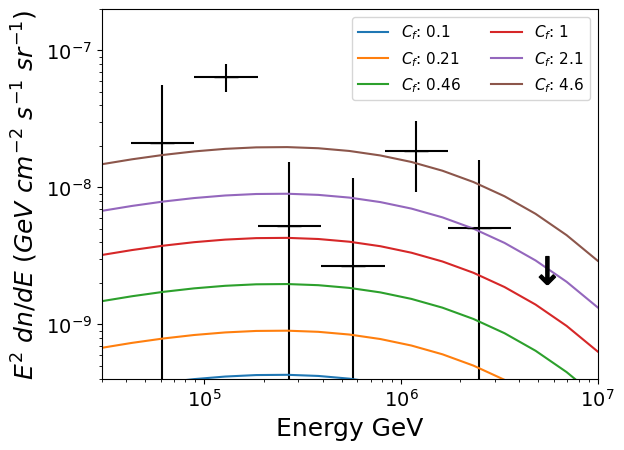}
    \caption{This figure shows the observed data from IceCube in black errorbar for data, the arrow for the upper limit, and the predicted diffuse neutrino flux from the outflow-cloud model in a different parameter of $C_{\rm f}=(\frac{\eta}{0.1})(\frac{C_{\rm v}}{0.1})(\frac{\dot{N}}{8\times 10^{-7}\rm Mpc^{-3}\ year^{-1}})(\frac{E_{\rm kin}}{10^{51}\rm erg})$ in 0.1 for the blue line, 0.21 for the orange line, 0.46 for the green line, 1 for the red line (the fiducial value), 2.1 for the purple line, and 4.6 for the brown line.}
    \label{Cf}
\end{figure}

\begin{figure}
    \centering
    \includegraphics[scale=0.49]{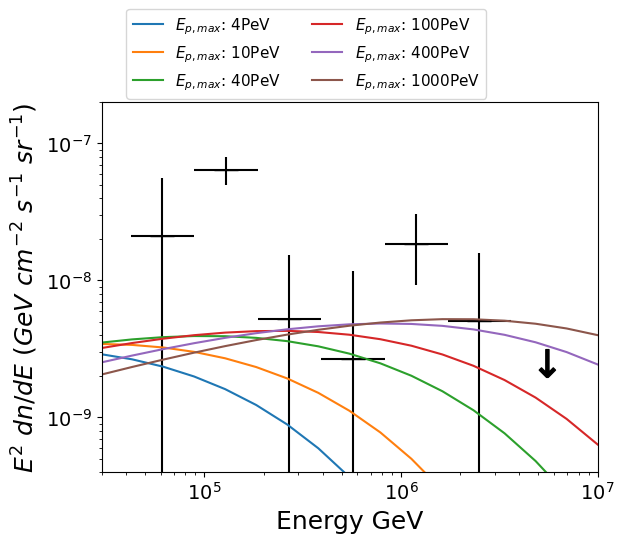}
    \caption{This figure shows the observed data from IceCube in black error bar for data, the arrow for the upper limit, and the predicted diffuse neutrino flux from the outflow-cloud model in different cutoff energy of the proton spectrum $E_{p,max}$ in 4 PeV for the blue line, 10 PeV for the orange line, 40 PeV for the green line, 100 PeV for the red line (the fiducial value), 400 PeV for the purple line, and 1000 PeV for the brown line.}
    \label{cutoff}
\end{figure}

To compare our predicted TDE diffuse neutrino flux with the observations, we have used neutrino flux data based on the latest research of diffuse neutrino observations from IceCube \citep{abbasi2021icecube} with the observation time $T_{o}$ 7.5 years \citep{abbasi2021icecube}, and data points are shown as the black error bars for the data and the arrows for the upper limit in Fig. \ref{index}, \ref{Cf}, \ref{cutoff}.

The fiducial values summarized in Tab. \ref{fidu} are used in our model and each parameter of the kinetic energy of outflows $E_{\rm kin}$, the cutoff energy of protons spectrum $E_{\rm p,max}$, the TDE volume rate $\dot{N}$, the index of protons spectrum $\Gamma$ is the average value on the distance. For the fiducial values, we tested the different indexes of the proton spectrum $\Gamma$, which is illustrated in Fig. \ref{index} (the fiducial index -1.7 for the green line). Due to the acceleration based on DSA \citep{drury1983introduction} and the outflow velocity reaching the order of $\sim 0.1\rm c$ \citep{malkov2001nonlinear}, we tested the range of index from -2 to -1.5 \citep{celli2020spectral} in a step size of 0.1, which is marked in the different colors illustrated in the legend of the figure. The harder spectrum has more diffuse neutrino flux and the index $\Gamma=-1.5$ is very close to the upper limit at 0.6 PeV. It is hard to distinguish the increase neutrino in the index $\Gamma$ from -1.7 to -1.5 below 0.5 PeV. At $\sim 0.1\rm\ PeV$, the increase in the index from -2 to -1.5 makes diffuse neutrino flux 4 times enhancement, and over PeV, the enhancement is nearly 10 times.

Due to the development of TDE search, much more TDEs have been found in recent years pushing more accuracy of TDE rate, thus, we test the different volume TDE rates with a range over an order of magnitude from $1\times 10^{-7}\rm Mpc^{-3}\ year^{-1}$ to $3\times 10^{-6}\rm Mpc^{-3}\ year^{-1}$, where the lowest value corresponds to the value from the early research \citep{donley2002large,esquej2008evolution,gezari2009luminous,van2014measurement} which is relatively low value because of the lack of sample of TDEs, and the highest value corresponds to the value from the theoretical research \citep{magorrian1999rates,wang2004revised} which is relatively high value because they estimate the effective galaxies of TDEs with the progress of M-$\sigma$ relation. The results of diffuse TDE rates are presented in Fig. \ref{Cf} marked in the different colors illustrated in the legend of the figure (the fiducial value $8\times 10^{-7}\rm Mpc^{-3}\ year^{-1}$ for the red line). As shown in Fig. \ref{Cf}, the volume TDE rate is proportional to the diffuse neutrino flux, then the IceCube observation data can limit the volume TDE rate based on the diffuse neutrino flux. 


The research on the energy of outflows is quite consistent no matter on numerical study \citep{lu2020self,curd2019grrmhd} or radio flare study \citep{alexander2016discovery,alexander2017radio,stein2021tidal,cendes2021radio,mou2022radio}, which suggest the typical energy at $1\times 10^{51}\rm erg$ and the highest energy the outflows could even reach $\sim 1\times 10^{52}\rm erg$. There also is quite an uncertainty that some researchers suggest $E_{\rm kin}\lesssim 1\times 10^{50}\rm erg$ for many TDEs \citep{matsumoto2021radio}. Here, we test the range of the energy of outflows $E_{\rm kin}$ from $1\times 10^{50}\rm erg$ to $3\times 10^{51}\rm erg$, the predicted diffuse neutrino fluxes are presented in Fig. \ref{Cf} marked in the different colors illustrated in the legend of the figure (the fiducial value $E_{\rm kin}=1\times 10^{51}\rm erg$ for the red line). As the energy of outflows increases, the diffuse neutrino flux would grow proportionally, and then the IceCube observation data can constrain the energy of outflows of an upper limit value $E_{\rm kin}\sim 1\times 10^{51}\rm erg$.


There is still uncertainty about the maximum energy of acceleration by DSA. In DSA, the square of outflow velocity $V_{o}$ is proportional to the cutoff energy of the protons spectrum $E_{\rm p,max}$ \citep{drury1983introduction,wu2022could}, thus, we test a large range from outflow velocity $V_{o}\sim 0.02c$ corresponding to the cutoff energy $E_{\rm p,max}\sim 4\rm\ PeV$ to outflow velocity $V_{o}\sim 0.3c$ corresponding to the cutoff energy $E_{\rm p,max}\sim 1000\rm\ PeV$, which is illustrated in Fig. \ref{cutoff} marked in the different colors illustrated in the legend of the figure (the fiducial value $100\rm\ PeV$ for the red line). As the Fig. \ref{cutoff} displays, the cutoff energy of the proton spectrum has the most influence on the diffuse neutrino flux over PeV, but at 0.1 PeV, it didn't show a significant difference, besides, the full range of the neutrino flux with the different cutoff energies is still within the limitation of the IceCube observation.

\begin{figure}
    \centering
    \includegraphics[scale=0.49]{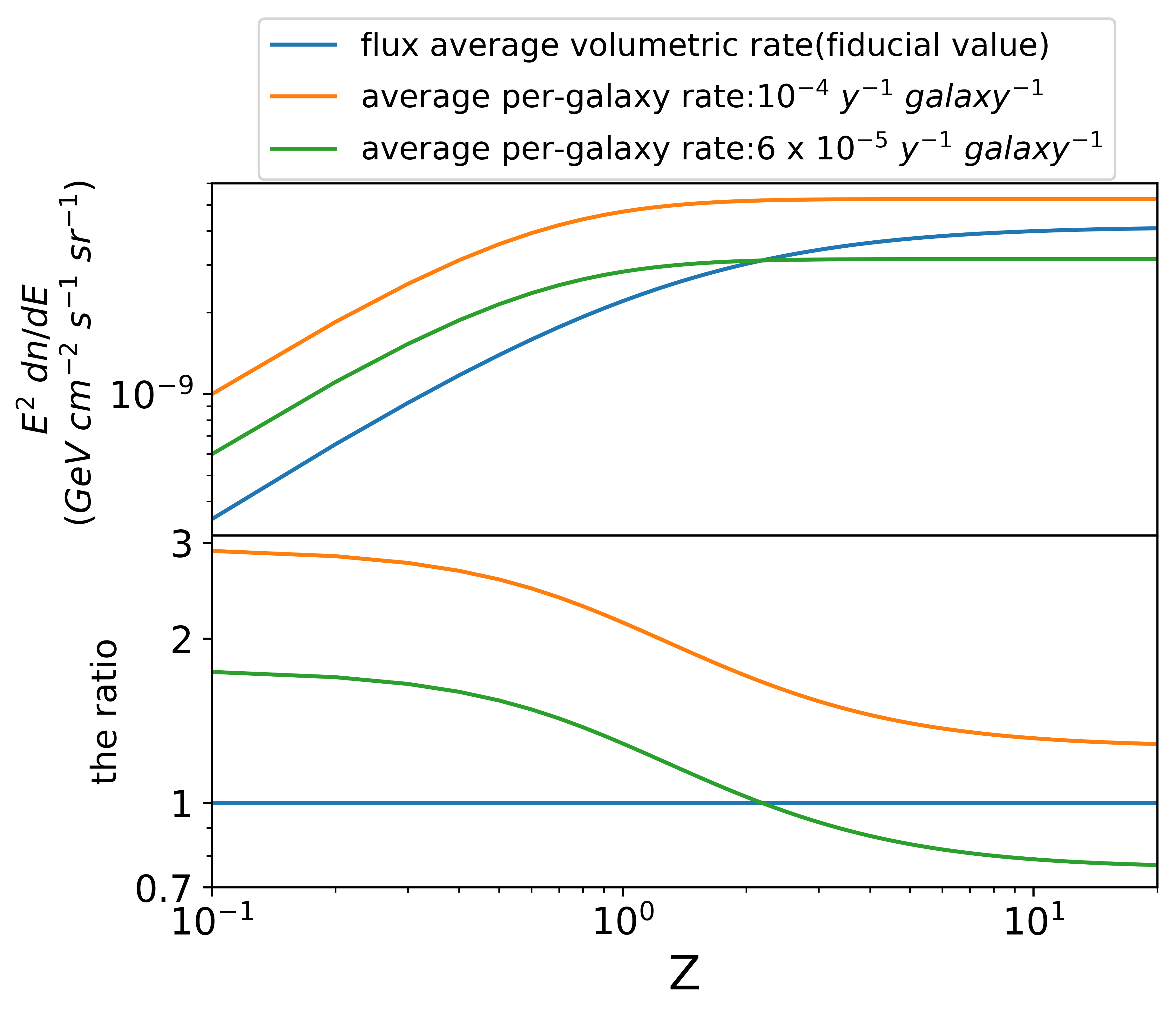}
    \caption{This figure shows the diffuse neutrino accumulation as a function of redshift $Z$ in the top panel, and the ratio in the bottom panel. The blue line represents the fiducial value, the yellow line for $\dot{n}=10^{-4}\rm\ year^{-1}\ galaxy^{-1}$, the green line for $\dot{n}=6\times 10^{-5}\rm\ year^{-1}\ galaxy^{-1}$.}
    \label{evolutioncom}
\end{figure}

\begin{figure}
    \centering
    \includegraphics[scale=0.49]{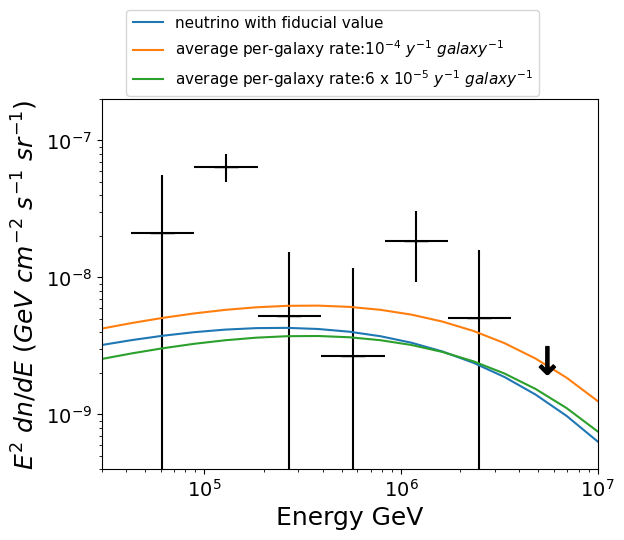}
    \caption{This figure shows the diffuse neutrino flux in fiducial value (blue line), $\dot{n}=10^{-4}\rm\ year^{-1}\ galaxy^{-1}$ (yellow line), and $\dot{n}=6\times 10^{-5}\ year^{-1}\rm\ galaxy^{-1}$ (green line).}
    \label{evolution}
\end{figure}

The TDE rate has been studied for tens of years \citep{gezari2021tidal}. However, up to now, the TDE rate is still under debate, furthermore, the evolution of the TDE rate as a function of redshift is also uncertain. So, we adopt a volume flux of the average TDE rate ($8\times 10^{-7}\rm Mpc^{-3}\ year^{-1}$) as the fiducial value, while the TDE occurring rate often could have an evolution with the redshift $Z$. Thus, we compare the results of the fiducial value with considering the evolution \citep{sun2015extragalactic} as a function of redshift $Z$:
\begin{equation}
    f_{\rm TDE}(Z)= [ (1+Z)^{-0.4}+(\frac{1+Z}{1.43})^{6.4}+(\frac{1+Z}{2.66})^{14.0}]^{-0.5}.
\end{equation}
We replace $\dot{N}$ as $\dot{n}\times N_{\rm gal,Z=0}\times f_{\rm TDE}(Z)$ in Eq. \ref{neuba}, where the $\dot{n}$ represents the per-galaxy TDE rate, the $N_{\rm gal,Z=0}$ for the number density of galaxy which can produce TDE at $Z=0$. We adopt two values: $\dot{n}=10^{-4}\rm\ year^{-1}\ galaxy^{-1}$ based on the results of the theoretical expectations \citep{magorrian1999rates,wang2004revised} and the observation estimation from host galaxy stellar mass function \citep{van2018mass,van2020optical}; the second one $\dot{n}=6\times 10^{-5}\rm\ year^{-1}\ galaxy^{-1}$ based on the results of the observation estimation from host galaxy black hole mass function \citep{van2018mass,van2020optical}. As for $N_{\rm gal,Z=0}$, we take $2.43\times 10^{-2}\rm Mpc^{-3}$ \citep{shankar2013accretion,ding2020mass,yang2021evolution}. The comparison of diffuse neutrino accumulation as a function of redshift $Z$ at 60 TeV is shown in Fig. \ref{evolutioncom} and the comparison of diffuse neutrino flux is shown in Fig. \ref{evolution}. For the total diffuse neutrinos, the prediction of $\dot{n}=10^{-4}\rm\ year^{-1}\ galaxy^{-1}$ is about 20\% higher than the fiducial value, and the result of $\dot{n}=6\times 10^{-5}\rm\ year^{-1}\ galaxy^{-1}$ is similar to the case of the fiducial value. With the evolution of the TDE rate, there is almost no contribution at $Z\geq 1$ instead of $Z\geq 2$ in fiducial value.

\section{conclusion and discussion}\label{results}
In this study, we have calculated the diffuse neutrino flux contributed by the TDEs based on the outflow-cloud model, which can significantly contribute to the diffuse neutrino detected by IceCube.
Our model predicts that the outflow-cloud interactions with fiducial parameter values can contribute significantly to the diffuse neutrino. Specifically, our model can account for approximately 80\% of the diffuse neutrino at energies near 0.3 PeV. Furthermore, below 0.1 PeV, our model still contributes approximately 18\% to the IceCube observed data. 
The pp interaction can also produce high-energy gamma rays. The extragalactic diffuse gamma-ray from Fermi-LAT \citep{ackermann2015spectrum} is a little higher than the diffuse neutrino flux and the diffuse gamma-ray concentrates below hundreds of GeV. The cascade emission from the $\gamma\gamma$ absorption by the extragalactic background light \citep{dwek2013extragalactic} and the CMB improve the gamma-ray flux by one order of magnitude in 10-100 GeV \citep{murase2013testing}. However, in our model, the gamma-ray flux is lower than the extragalactic diffuse gamma-ray from Fermi-LAT \citep{ackermann2015spectrum} by over two orders of magnitude, which can't limit our model.

We also explored the impact of varying each parameter. The index of protons spectrum $\Gamma$ changes from -2 to -1.5, leading to the diffuse neutrino flux enhancement about 4 times at around 0.1 PeV as well as near 10 times over PeV as illustrated in Fig. \ref{index}. For the cutoff energy of protons spectrum $E_{\rm p,max}$ from 4 PeV to 1000 PeV, as illustrated in Fig. \ref{cutoff}, the most influence from the cutoff energy of protons spectrum $E_{\rm p,max}$ is the maximum energy of the diffuse neutrino flux and has the little influence on the diffuse neutrino flux blow 0.4 PeV and over 200 PeV, the diffuse neutrino flux predicted by our model would exceed the IceCube observation data around 3.5 PeV band. Because there are some parameters ($\eta$, $C_{\rm v}$, $\dot{N}$, $E_{\rm kin}$) which are directly proportional to the changes in the diffuse neutrino flux, thus, we set a parameter $C_{\rm f}=(\frac{\eta}{0.1})(\frac{C_{\rm v}}{0.1})(\frac{\dot{N}}{8\times 10^{-7}\rm Mpc^{-3}\ year^{-1}})(\frac{E_{\rm kin}}{10^{51}\rm erg})$. $C_{\rm f}=1$ is the fiducial value and if with $\Gamma=-1.7$, our model prefers a $C_{\rm f}$ less than 2, the results of which are shown in Fig. \ref{Cf}.

\begin{figure}
    \centering
    \includegraphics[scale=0.49]{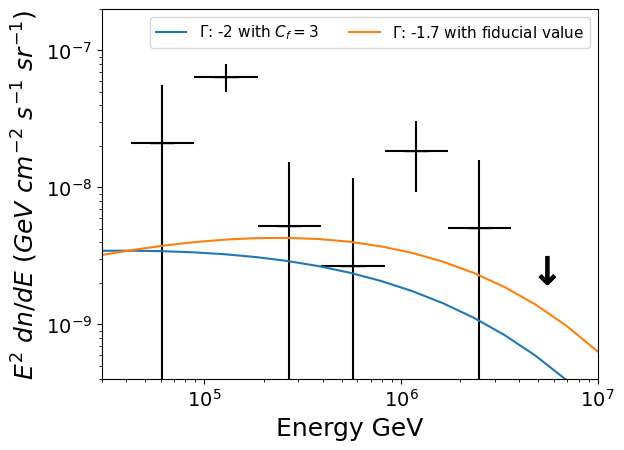}
    \caption{This figure shows the observed data from IceCube in black errorbar for data, the arrow for the upper limit, and the predicted diffuse neutrino flux from the outflow-cloud model in different parameters of $C_{\rm f}$ in 3 and the $\Gamma=-2$ for the blue line, and the fiducial value for the orange line.}
    \label{Cf3}
\end{figure}

In the standard theory of non-linear DSA, the total compression ratio could be over 7 and accompanied by the particle feedback on the shock producing harder spectra which lead to $\Gamma\lesssim -1.5$ \citep{malkov2001nonlinear}, thus, we chose the upper range of $\Gamma$ as -1.5 \citep{celli2020spectral}. With the observation of supernovae and supernova remnant \citep{chevalier2006circumstellar,morlino2012strong}, the spectra index $\Gamma$ would be around -2 \citep{evoli2019galactic,evoli2020ams}, then, a revised DSA theory explains that the sub shocks accelerating particle can provide a softer spectrum ($\Gamma\sim -2$) with a large faction ($\eta\sim 0.3$) \citep{caprioli2012cosmic,haggerty2020kinetic,caprioli2020kinetic}. Thus, we compared the results of the revised DSA theory \citep{caprioli2012cosmic,haggerty2020kinetic,caprioli2020kinetic} (except $\eta$ and $\Gamma$, other parameters at fiducial values) in Fig. \ref{Cf3}, which showed that above 0.1 PeV, the flux with $\Gamma$=-2, $\eta=0.3$ is lower than the fiducial value by factor 2 around 1 PeV, and softer CR spectra will accommodate a higher value of $C_{\rm f}$.

\begin{figure}
    \centering
    \includegraphics[scale=0.49]{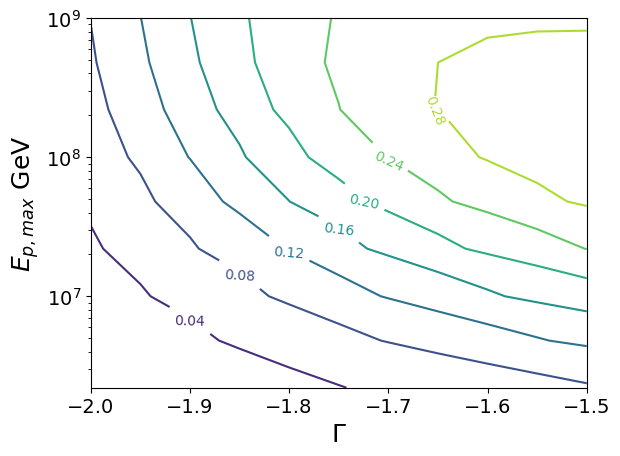}
    \caption{This contour figure shows the total contribution fraction depending on the combination of the cutoff energy $E_{\rm p,max}$ (Y-axis) and the index $\Gamma$ (X-axis) of the proton spectrum.}
    \label{contour}
\end{figure}

\begin{figure}
    \centering
    \includegraphics[scale=0.49]{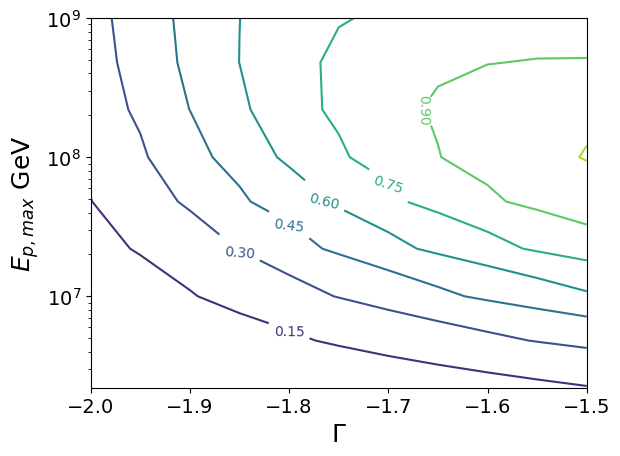}
    \caption{This contour figure shows the contribution fraction in the energy range from 0.19-0.39 PeV depending on the combination of the cutoff energy $E_{\rm p,max}$ (Y-axis) and the index $\Gamma$ (X-axis) of the proton spectrum.}
    \label{300contour}
\end{figure}

The total fraction of IceCube's neutrino flux is also considered, including the variation of every parameter described in the above paragraphs. We calculate the total fraction by the ratio between the total neutrino flux we predict and the IceCube data \citep{abbasi2021icecube}, and the uncertainty of the IceCube data gives the error range. Firstly, based on the fiducial values, there is $\sim 24^{+2}_{-15}\%$ total contribution in the total IceCube's neutrino flux, which is consistent with the stacking analyses for TDEs \citep{stein2019search}. For the cutoff energy of proton spectrum $E_{\rm p,max}$ varying from 4 PeV to 1000 PeV, the total contribution changes from $7^{+1}_{-4}\%$ to $26^{+2}_{-17}\%$. Due to the constraint from the stacking research \citep{stein2019search,bartos2021icecube}, $C_{f}$ should be less than 1.3 if we limit the total contribution in $\sim 30\%$\citep{stein2019search,bartos2021icecube}. As for the index of proton spectrum $\Gamma$ varying from -2 to -1.5, the total contribution changes from $8^{+1}_{-5}\%$ to $27^{+3}_{-17}\%$. Due to the index $\Gamma$ and the cutoff energy $E_{\rm p,max}$ of the proton spectrum is degenerate, the contribution in the energy range from $\sim 0.19-0.39$ PeV of these two parameters is explored and the results are shown in Fig. \ref{300contour} and the total contribution is shown in Fig. \ref{contour}.

Compared to the possible high energy diffuse neutrino contributed by blazars \citep{padovani2015simplified}, first of all, the results presented in Figs. \ref{index}, \ref{Cf}, \ref{cutoff} have the same units on the vertical axis with the results (their Figures 3 \& 4 in \citep{padovani2015simplified}) of the blazer diffuse neutrino model. In addition, the IceCube data used in the blazer model and our work have a negligible difference because they use the data published in 2014 \citep{aartsen2014observation} and we use the data published in 2021 \citep{abbasi2021icecube}. In the blazer model, the most contribution of the diffuse neutrino over 10 PeV is produced by the high energy peaked blazers (HBL), however, at sub-PeV the contribution down to $\sim 10\%$ (could increase to $\sim 20\%$ for some parameters). In contrast, our outflow-cloud model predicts the most substantial contribution around 0.3 PeV which can reach 80\% and below 0.1 PeV, there are still 18\% contributions above PeV, and our model would have a cut-off at tens of PeV. Both studies suggest relatively low contributions around 0.1 PeV, indicating the possibility of another mechanism responsible for producing neutrinos in that energy range below 0.1 PeV.


\begin{acknowledgments}
We thank the anonymous referee for the thoughtful comments and suggestions. This work is supported by the National Key Research and Development Program of China (Grants No. 2021YFA0718503 and 2023YFA1607901), the NSFC (12133007,12003007), and the Fundamental Research Funds for the Central Universities (No. 2020kfyXJJS039).
\end{acknowledgments}
\bibliographystyle{apsrev4-2}
\bibliography{apssamp.bib}

\end{document}